\documentclass[superscriptaddress,twocolumn,showkeys,footinbib,amssymb,aps]{revtex4}

\usepackage{graphicx,dcolumn,bm,amsmath,color,bm}

\newcommand{\eq}[1]{Equation~(\ref{#1})}

\newcommand{\fig}[1]{Figure~\ref{#1}}

\newcommand{\eeq}{ \end{equation} }
\newcommand{\beq}{ \begin{equation} }

\newcommand{\eea}{ \end{align} }
\newcommand{\bea}{ \begin{align} }

\newcommand{\bhua}{ {\bf \hat{u}}_{1} }
\newcommand{\bhub}{ {\bf \hat{u}}_{2} }

\newcommand{\bhu}{ {\bf \hat{u}} }

\newcommand{\bfr}{ {\bf r} }

\newcommand{\bn}{ {\bf \hat{n}} }
\newcommand{\kbt}{k_{B}T}
\newcommand{\ellp}{\ell^{\prime}}

\newcommand{\talpha}{\tilde{\alpha}}

\begin{document}

\title{Effect of size polydispersity on the pitch of nanorod cholesterics}
\author{H. H. Wensink}
\email{wensink@lps.u-psud.fr}
\affiliation{Laboratoire de Physique des Solides - UMR 8502, CNRS, Universit\'e Paris-Sud, Universit\'{e} Paris-Saclay, 91405 Orsay, France}
\date{\today}

\begin{abstract}

Many nanoparticle-based chiral liquid crystals are composed of polydisperse  rod-shaped particles with considerable spread in size or shape, affecting the mesoscale chiral properties in, as yet, unknown ways. Using an algebraic interpretation of Onsager-Straley theory for twisted nematics, we investigate the role of length polydispersity   on the pitch of nanorod-based cholesterics  with a continuous length polydispersity, and find that polydispersity enhances the twist elastic {modulus, $K_{2}$,} of the cholesteric material without affecting the {effective helical amplitude, $K_{t}$}. In addition, for the infinitely large average aspect ratios considered here, the dependence of the pitch on the overall rod concentration is completely unaffected by polydispersity.  For a given concentration, the increase in twist elastic {modulus} (and reduction of the helical twist) may be up to 50\% for strong size polydispersity, irrespective of the shape of the unimodal length distribution. We also demonstrate that the twist reduction is reinforced in bimodal distributions, by doping a polydisperse cholesteric with very long rods.  Finally, we identify a subtle, non-monotonic change of the pitch across the isotropic-cholesteric biphasic region.

\end{abstract}

\maketitle

\section{Introduction}

Polydispersity is widespread in colloidal and polymeric systems, since  the building blocks are never fully identical but exhibit a continuous spread in size, shape, or surface charge \cite{ FASOLO2004}.
The variety in microscopic interactions ensuing from polydispersity may have a considerable influence on the phase stability \cite{huang_1991,sollichoverview} or the mechanical properties of nanoparticle-based materials through aggregation \cite{bushell_jcis1998}, packing~\cite{farr_jcp2009}, or percolation processes  \cite{kyrylyuk_PNAS2008}. Research efforts can be aimed at either purifying colloidal suspensions in order to promote crystallization \cite{murray_kagan}, such as through templating \cite{blaaderen_wiltzius1997}, or at deliberately enhancing size polydispersity; for instance, to improve the electronic conductivity of percolated rod \mbox{networks~\cite{kyrylyuk_NatNANO2011,kyrylyuk_PNAS2008}}, to stabilize glassy states of spherical particles \cite{auer_nature2001,pusey_2009}, or to realize complex fluids  with bespoke rheological properties \cite{tenbrinke_sm2008}.  The effect of size polydispersity in lyotropic liquid crystals composed of non-spherical (e.g., rod-shaped) nanoparticles was first addressed in the 1980s, focussing mostly on its effect on the nematic osmotic pressure  \cite{odijkgauss}, on the stability of smectic order \cite{sluckin1989}, and on the impact of size bidispersity on the order-disorder transition \cite{OdijkLekkerkerker, Vroege92}. 

The presence of chiral forces among rod-shaped particles is usually expressed in terms of some helical organization on the mesoscale, as is the case, for instance, in chiral nematics or cholesterics~\cite{gennes-prost,dierking}. The helical twist of the local nematic director defines a typical mesoscopic lengthscale, referred to as the pitch, whose controllability is of key importance in the manifold examples of chiral nematics involved in technological applications (e.g., displays), as well as in nature \cite{mitov2017}.
Cholesteric materials based on nanorods  commonly consist of rigid, fibrillar units, composed of some biological component such as cellulose (CNCs) \cite{gray-cullulose, heux2000,lagerwall2014a},  chitin \cite{REVOL1993,belamie2004}, collagen \cite{giraud2008a}, or amyloid \cite{nystrom2018,bagnani2019}.
These fibrils are inherently size-polydisperse and the effect of size disparity on the sensitivity of the pitch remains an important outstanding issue. {In these systems,  size polydispersity is {quenched} by the synthesis procedure and usually does not depend on the thermodynamic state of the system. Similar to chiral chromonics \cite{chromo2008},   nanometric chiral building blocks, such as short-fragment DNA \cite{zanchetta2010a,demichele2016},  may reversibly  polymerize into chiral filaments that are inherently polydisperse. However, these systems constitute a different class of cholesterics,  characterized by {annealed} polydispersity where the contour length distribution of the filaments is dictated by temperature, the degree of semiflexibility, and the monomer concentration~\cite{Taylor_1993,schoot1994}.}

In this paper, we attempt to address the effect of {quenched length polydispersity}  on cholesterics from a theoretical viewpoint, and propose  an algebraic theory that is capable of linking the cholesteric pitch to the microscopic chirality of the rods, as well as their inherent length distribution. We find that length-polydispersity has a significant impact on the twist elastic {modulus} of the cholesteric material, increasing it by about 50\% compared to its monodisperse counterpart at the same overall rod concentration. Within the same framework, we also address the isotropic-cholesteric phase coexistence and identify the concentration, length-composition, and pitch  of the cholesteric phase fraction upon traversing the biphasic {region}, revealing subtle non-monotonic trends that could be exploited to purify or control the size composition of a cholesteric material.  We hope that the present theory may serve as a useful tool in guiding or rationalizing certain experimental trends regarding the pitch of  biofibrillar-based cholesteric systems systems with quenched length polydispersity.

\section{Onsager-Straley Theory for Polydisperse Cholesterics}
 Let us start with the free energy per unit volume $V$ of a polydisperse assembly of strongly elongated rods with diameter $D$ and length $L$, the latter following some quenched length distribution $c(\ell)$ with renormalized rod length $\ell$. Within Onsager's second-virial approximation \cite{Onsager},  the free energy of the rod fluid per unit volume  reads:
\begin{align}
f = \frac{v_{0}F}{ V}  \sim & \int d \ell c(\ell) (\ln c(\ell) - 1) + \int d \ell c(\ell) \sigma^{(\ell)}   \nonumber \\ 
& + \iint d \ell  d \ellp c(\ell)  c(\ellp) \left [ \rho^{(\ell \ellp )}  + f_{c}^{(\ell \ellp)}(q) \right ],
 \label{free}
\end{align}
where $\beta = (\kbt)^{-1}$ denotes the thermal energy in terms of Boltzmann's constant $k_{B}$ and temperature $T$.  The renormalized rod length $\ell = L/L_{0}$ exhibits a continuous spread prescribed by a normalised distribution $p(\ell)$, so that $c(\ell) = c_{0} p(\ell)$ in terms of the overall dimensionless particle density \mbox{$c_{0} = Nv_{0}/V $} and microscopic volume $v_{0} = \pi L_{0}^{2} D/4$, with  $L_{0}$ the {average} rod length. Consequently, the first moment of the distribution is fixed at unity (i.e.,  $\int d \ell p(\ell) \ell =1$).

The free energy consists of  three entropic contributions relating to the ideal gas, orientational, and excluded volume entropy, respectively.  The first two entropic quantities can be computed in their exact form, while the excluded-volume entropy is defined on the level of the second-virial coefficient between a pair of rods. This approximation should be accurate 
if all rod species are sufficiently slender and that their aspect ratio $L/D \gg 1$ \cite{Onsager}.  The orientational entropy is defined as:
\begin{align}
\sigma^{(\ell)} & =\int d \Omega \psi(\Omega, \ell) [ 4 \pi \psi ( \Omega, \ell ) ],
\label{sigma}
\end{align}
and involves some unknown orientational distribution function $\psi(\Omega, \ell)$  that describes the orientational probability of a rod with length $\ell$ in terms of a solid angle $\Omega$.  Trivially, for an isotropic fluid, where the rods point in random directions, the distribution becomes a mere constant  $\psi  = (4 \pi)^{-1}$, irrespective of $\ell$. The orientational entropic factor is then simply rendered zero (i.e.,  $\sigma = 0$).

\begin{figure}
\centering
\includegraphics[width= 0.9 \columnwidth]{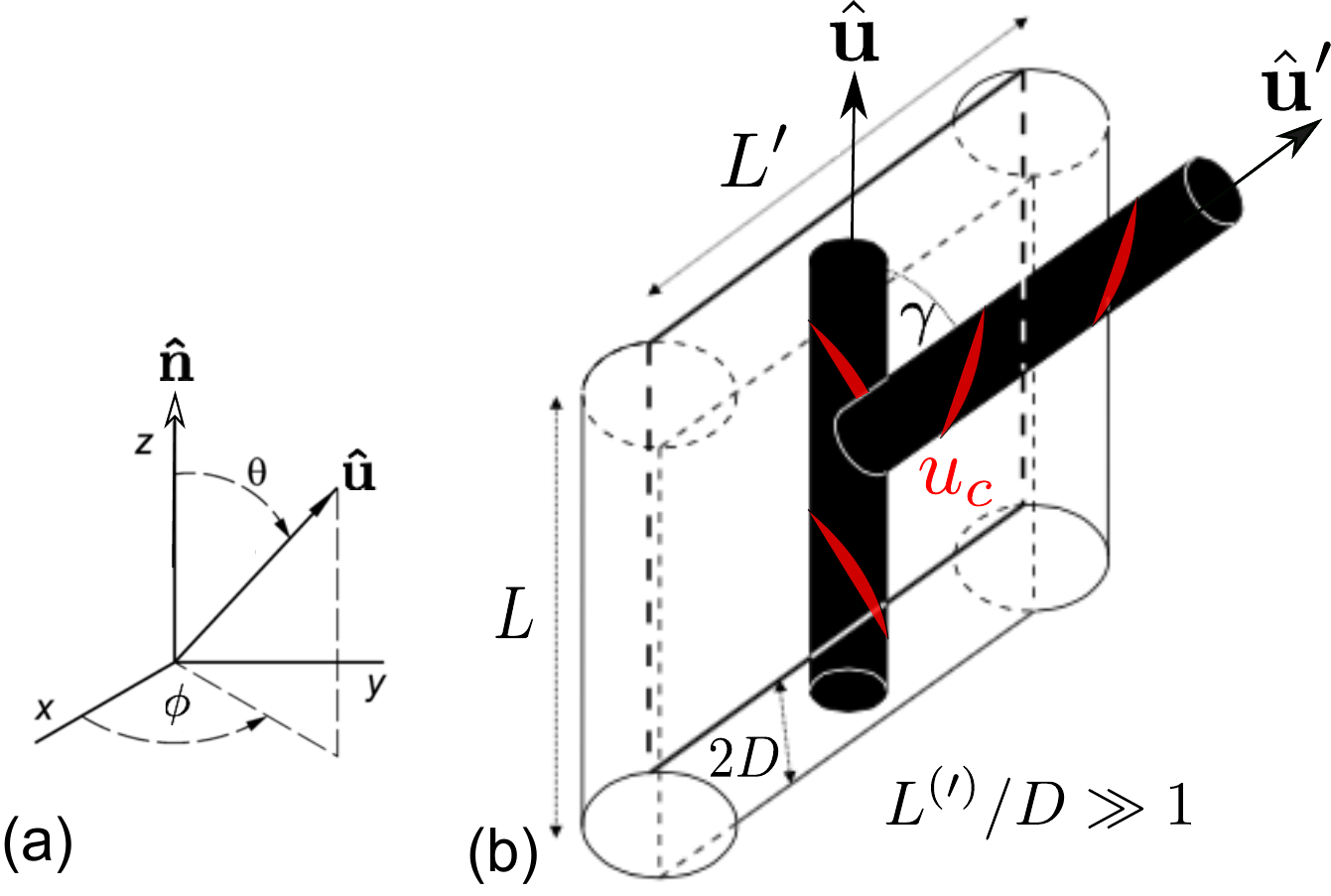}
\caption{ \label{fig1} {(\textbf{a}) Overview of the lab frame, nematic director $\bn$, and principal angles used in the present analysis; (\textbf{b}) Excluded volume between two achiral hard cylinders decorated with a perturbative chiral potential  $u_{c}$ acting locally along the rod contour (indicated by the red helical threads).  The rod excluded volume is assumed to be unaffected by the chiral potential and is responsible for stabilizing the nematic order ($\rho$) and generating twist elasticity ($K_{2}$), while the chiral potential $u_{c}$ promotes director twist ($K_{t}$). } }
\end{figure}

The second entropic contribution  $\rho$ is defined as the angular-averaged excluded volume per particle in a nematic phase, normalized to its random isotropic average:
\beq
\rho^{(\ell, \ellp )}    = \frac{4} {\pi} \ell \ellp  \iint d \Omega d \Omega^{\prime}  \psi(\Omega, \ell) \psi(\Omega^{\prime}, \ellp)  | \sin \gamma |,  
\label{rho}
\eeq
with $\gamma$ denoting the enclosed angle between two rods ({see \fig{fig1}}). For the isotropic phase, it is easily established that $ \langle \langle  | \sin \gamma |  \rangle \rangle_{\psi} =\pi/4$ and $\rho^{(\ell \ellp)} =\ell \ellp$. 

The last contribution in \eq{free}  is due to Straley \cite{straleychiral}, and describes the {free energy difference between the weakly twisted director field  of a cholesteric liquid crystal and the uniform  one of a nematic}.  The degree of helical organization is defined in terms of a dimensionless {wave number,  $q = 2 \pi L_{0} /p_{c}$, where the pitch is required to be much larger than the average nanoparticle size (i.e., $p_{c} \gg L_{0}$)}.  Under these restrictions, $q \ll 1$, and the additional free energy density takes a simple quadratic form:
\beq
f_{c} ^{(\ell \ellp)} (q) =   q K_{t}^{(\ell \ellp)} + \frac{1}{2} q^{2} K_{2}^{(\ell \ellp)},  
\eeq
in terms of a species-dependent helical amplitude $K_{t}^{(\ell \ellp)}$ and twist elastic modulus  $K_{2}^{(\ell \ellp)}$, defined microscopically as:
\begin{align}
K_{t}^{(\ell \ellp)} & \sim   \iint d \Omega d \Omega^{\prime}  \psi (\Omega, \ell)  \dot{\psi} (\Omega^{\prime}, \ellp) \Omega^{\prime}_{\perp} M_{t}^{(\ell \ellp)}(\Omega, \Omega^{\prime}),   \nonumber \\
K_{2}^{(\ell \ellp)} & \sim  \iint d \Omega d \Omega^{\prime}  \dot{\psi} (\Omega, \ell)  \dot{\psi} (\Omega^{\prime}, \ellp) \Omega_{\perp}  \Omega^{\prime}_{\perp}  M_{2}^{(\ell \ellp)}(\Omega, \Omega^{\prime}).
\label{kaas} 
\end{align}
These expressions depend on the derivative of the local orientational probability  $\dot{\psi} =  \partial \psi / \partial \Omega $ with $\Omega_{\perp}$ denoting the component of the rod orientation perpendicular to the local nematic director and the pitch axis.  The kernels describe the interactions between the chiral rods, which we assume to consist of a weak soft potential $u_{c}$ imparting chirality {superimposed onto a hard-core repulsion generated by the cylindrical  backbone that is responsible for generating twist elasticity.} The helical amplitude  $M_{t}$ is given by an integrated (van der Waals) potential \cite{varga2006a,wensinkjackson}:   
\beq
M_{t}^{(\ell \ellp)}(\Omega, \Omega^{\prime}) \sim (v_{0} L_{0})^{-1}   \int_{\notin v_{\rm excl}} d \bfr \bfr^{\parallel} \beta u_{c}^{(\ell \ellp)} (\bfr, \Omega, \Omega^{\prime} ),
\label{mt}
\eeq
{and depends uniquely on the  chiral potential $u_{c}^{(\ell \ellp)}$} between rods of length $\ell$ and $\ellp$, which we will specify shortly. Here, $\bfr_{\parallel}$ represents the component of the centre-of-mass distance between a rod pair along the pitch axis. The second kernel, $M_{2}$, depends on a generalized excluded-volume between the achiral  cylindrical backbone of two rods of different lengths, and reads for slender rods \cite{odijkchiral,wensinkjackson}:
\beq
M_{2} ^{(\ell \ellp)}(\Omega, \Omega^{\prime}) \sim -\frac{2}{3 \pi} \ell \ellp | \sin \gamma | (\ell ^{2} \Omega_{\parallel}^{2} + \ell^{\prime 2} \Omega^{\prime 2}_{\parallel}  ),
\label{m2}
\eeq
where  $\Omega_{\parallel}$ is the rod orientation projected along the pitch axis.

Although the twisting of the director changes the local uniaxial alignment in favour of biaxial order \cite{harris-rmp}, the biaxial perturbation is very weak for $q \ll 1$, and we shall assume that the local uniaxial nematic order remains unperturbed.  Consequently, the orientational distribution depends solely on the polar angle, $\theta$,  between the main particle orientation vector, $\bhu$, and the nematic director, $\bn$, by $\cos \theta = \bhu \cdot \bn$.  Let~us further assume strongly  nematic order,  so that the use of a  Gaussian Ansatz \cite{odijkgauss,odijkoverview} for the local orientational probability is justified:
\beq
 \psi_{G} (\theta, \ell) \sim \frac{\alpha(\ell)}{4 \pi}  \exp \left (  -\frac{1}{2} \alpha(\ell) \theta^{2}  \right ),
 \label{gauss}
 \eeq
supplemented with its polar mirror form $\psi (\pi - \theta , \ell)$ along $-\bn$, in order to guarantee local apolar order. The variational parameter  $\alpha (\ell) $ is required to be much larger than unity and is length-dependent.  While $\alpha(\ell )$ is, as yet, unknown in explicit form, common sense tells us that $\alpha(\ell)$ should be proportional to the rod contour length, since long rods tend to be more strongly aligned than short rods \cite{OdijkLekkerkerker, odijkgauss}. The Gaussian approximation cannot represent isotropic order since, upon taking $\alpha \downarrow 0$, the expression above reduces to zero, rather than giving the desired form $\psi = 1/4\pi$.   There are  consistent algebraic trial functions for $\psi$  that do correctly render isotropic order in this limit, but these  involve more complicated distributions that tend to compromise the tractability of the theory \cite{Onsager, francomelgar2008}. 

From \eq{gauss} we readily find an asymptotic expression for the orientational entropic factor: 
\begin{align}
\sigma^{(\ell)} &  \sim  \ln \alpha(\ell) - 1,
\label{sigma_as}
\end{align}
whereas the excluded-volume term can be estimated from an asymptotic expansion for $\alpha \gg 1$ giving up to the leading order \cite{odijkoverview}:
\beq
\rho^{(\ell, \ellp )}    \sim \frac{4} {\pi} \ell \ellp  \left ( \frac{\pi}{2}  \right )^{\frac{1}{2}}  \left ( \frac{1}{\alpha(\ell)} + \frac{1}{\alpha(\ellp)}  \right )^{\frac{1}{2}}.
\label{rho_as}
\eeq
Since the effective torque associated with director twist is relatively weak compared to the one enforcing nematic order, it is safe to assume that $\alpha(\ell)$  does not depend on the pitch.  
Using the results of Equations~(\ref{sigma}) and (\ref{rho}) in the free energy of the untwisted nematic, \eq{free} (with $q=0$) enables a formal minimization with respect to $\alpha(\ell)$, giving the following  self-consistency condition:
\beq
\talpha^{\frac{1}{2}}(\ell) =2^{\frac{1}{2}} \int d \ellp \ell \ellp p(\ellp) g_{0}(\ell, \ellp),
\label{alfa}
\eeq
with
\beq
g_{0}(\ell, \ellp) = \left  ( 1 + \frac{\talpha(\ell)}{\talpha(\ellp)} \right )^{-\frac{1}{2}}.
\eeq
 No matter what length distribution, $\alpha(\ell)$ scales quadratically with concentration $c_{0}$, so that it is expedient to factorize 
\beq
\alpha (\ell) = \frac{4}{\pi} c_{0}^{2} \talpha (\ell),
\eeq
with $\talpha(\ell)$ depending only on the {shape} of the normalised distribution $p(\ell)$. Unfortunately, \eq{alfa} does not permit $\alpha(\ell)$ to be resolved analytically, but a numerical solution is easily obtained for a given distribution $p(\ell)$ \cite{wensink_jcp2003}. 

Minimizing \eq{free} with respect to $q$, we obtain 
the equilibrium value for the wave number $q$ reflecting a balance between the {helical} amplitude and twist elastic {modulus}:  
\beq
q \equiv \frac{K_{t}}{K_{2}} = - \frac{ \iint d \ell  d \ellp c(\ell)  c(\ellp) K_{t}^{(\ell \ellp)} }{ \iint d \ell  d \ellp c(\ell)  c(\ellp)  K_{2}^{(\ell \ellp)} }.
\label{eqpitch}
\eeq
These contributions will be computed in algebraic form in the next Section. 
 
\section{Asymptotic Results for the Helical Amplitude and Twist Elastic Modulus}

\begin{figure*}
\centering
\includegraphics[width= 1.6 \columnwidth]{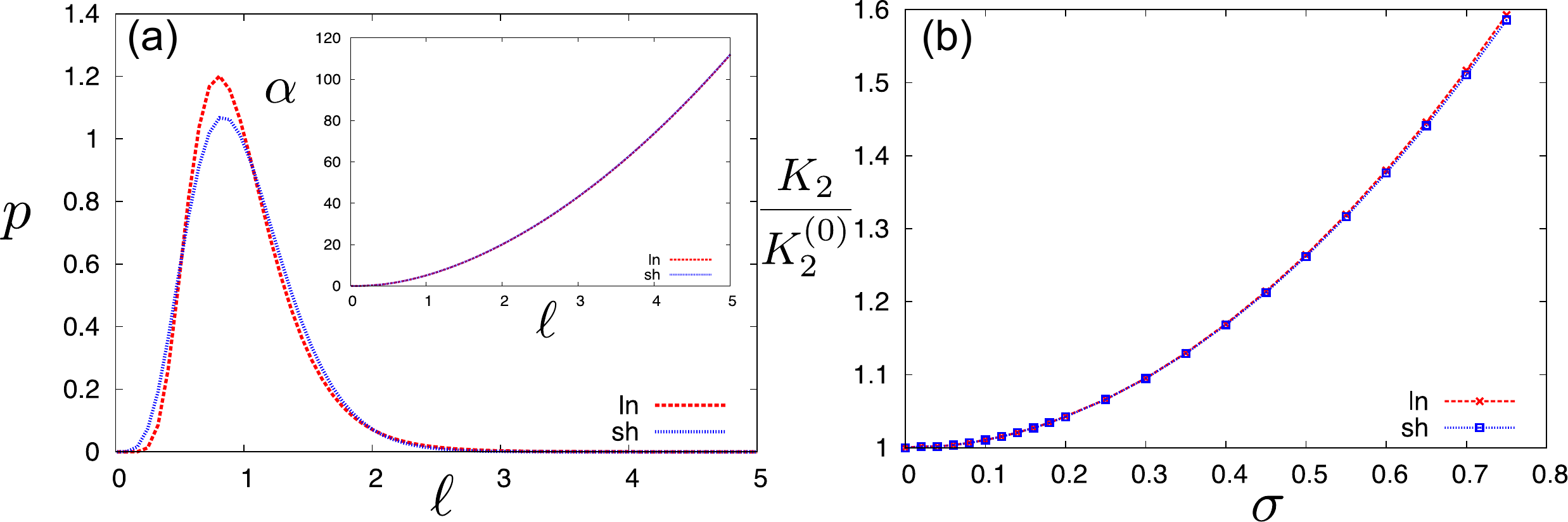}
\caption{ \label{fig2} (\textbf{a}) Overview of log-normal and Schulz-length distributions with polydispersity $\sigma = 0.4$. The~inset depicts the associated Gaussian parameter $\alpha$ versus rod length $\ell$, obtained from \mbox{\eq{alfa}}, showing that long rods align much more strongly than short ones; (\textbf{b})  Twist elastic modulus $K_{2}$ for a polydisperse nanorod cholesteric with increasing polydispersity $\sigma$, normalized to the value  $K_{2}^{(0)}$ of the corresponding monodisperse system.   The results for both distributions are virtually indistinguishable.}
\end{figure*}

Let us now  propose a simple  chiral potential acting between two freely rotating rods.  We shall  consider the commonly used pseudo-scalar form \cite{goossens,vargachiral}:
\beq
u_{c}^{(\ell \ellp)} (\bfr, \Omega, \Omega^{\prime} ) \sim \varepsilon g \left ( r \right ) ( \bhua \times \bhub \cdot {\hat{\bf r}} ),   
\eeq
with $g(r)$ some rapidly decaying function of the centre-of-mass distance $r$ and $\varepsilon$ specifying the microscopic chiral strength between the rods. We may work out the integrated chiral potential $M_{t}$ corresponding to this potential,  first by defining the pitch axis of the cholesteric to align along the $x-$axis of a Cartesian laboratory frame {(see \fig{fig1})}, and defining a rod orientation  $\bhu = (  \sin \theta \sin \varphi, \sin \theta \cos \varphi, \cos \theta ) $  in terms of polar and azimuthal angles $(\theta, \varphi)$, with respect to a reference nematic director $\bn$ pointing along the $z-$axis. Then, $\Omega_{\perp}  = u_{y}$  and we may perform a Taylor expansion for $\theta \ll 1$ and keep only the leading order contribution. Some algebraic manipulations, along the lines proposed in \cite{odijkelastic, odijkchiral}, lead to the following asymptotic expression: 
\beq
 \Omega^{\prime}_{\perp} M_{t}^{(\ell \ellp)} \sim \bar{\epsilon} \ell \ellp \left [ (\theta^{\prime 2} - \theta^{2} ) + |\gamma | ^{2}  \right ],
\eeq
with $ \bar{\epsilon}$ a dimensionless chiral strength combining various microscopic features:
\beq
\bar{\epsilon} \sim \frac{1}{\pi} \beta \varepsilon  \frac{D}{L_{0}} \int_{D}^{\infty} dr rg(r).
\eeq
A similar analysis can be performed for the twist elastic contribution $M_{2}$ producing the following angular dependency for strong alignment  \cite{odijkelastic}:
\begin{align}
 \Omega_{\perp} \Omega^{\prime}_{\perp} M_{2}^{(\ell \ellp)}   \sim &   -\frac{\ell \ellp}{24 \pi}  \left [ | \gamma | (\theta^{\prime 2} + \theta^{2} )(\ell^{\prime 2} \theta^{\prime 2} + \ell^{2} \theta^{2})   \right . \nonumber \\
 & \left .  - |\gamma | ^{3} (\ell^{\prime 2} \theta^{\prime 2} + \ell^{2} \theta^{2}) \right ]. 
\end{align}
The remaining task is to perform Gaussian orientational averages of these quantities, as per \eq{kaas}, to arrive at an explicit expression for the kernels $M_{t}$ and $M_{2}$. The mathematical theorem that allows one to compute the angular averages has been discussed in Onsager's original paper \cite{Onsager}, and used later on in Odijk's work on elastic constants \cite{odijkelastic}. The averages are given in explicit form in Appendix A.  An~additional advantage of the Gaussian approach is that we can use the simple relation $\dot{\psi}_{G} \sim \alpha(\ell) \psi_{G}  $ to  obtain the derivate of the orientational distributions involved in \eq{kaas}.  Straightforward algebraic manipulation then leads to a simple result for the helical amplitude:
\beq
K_{t} ^{(\ell \ellp)} \sim  2 \bar{\epsilon} \ell \ellp.  
\label{ktll}
\eeq 
The overall helical amplitude is {\em independent} of the length distribution, and scales quadratically with rod concentration $c_{0}$:
\beq
K_{t} \sim 2 c_{0}^{2}  \bar{\epsilon}.
\eeq
We remark that this result may be different for purely steric chirality induced by some helical nanorod shape, such as a corkscrew \cite{dussi2015a,kolli2014b,morales2015}. {More complicated chiral interactions---for example, those generated by a helical arrangement of charged surface groups, as in the case of viral rods \cite{tombo2006}---can, in principle, be~captured within a numerical interpretation of the van der Waals term \eq{mt}.}
The twist elastic modulus for a polydisperse nematic takes on  a more elaborate form:
\begin{align}
K_{2}     \sim & \frac{c_{0}  }{12 \pi 2^{\frac{1}{2}} } \iint d \ell  d \ellp p(\ell  )  p (\ellp)  \ell \ellp  \left ( \frac{1}{\tilde{\alpha}(\ell) } + \frac{1}{\tilde{\alpha}(\ellp)} \right )^{\frac{1}{2}}  \nonumber \\
& \times  \frac{\ell^{2} \tilde{\alpha}(\ellp) [4 \tilde{\alpha}(\ell) + 3 \tilde{\alpha}(\ellp) ] + \ell^{\prime 2} \tilde{ \alpha}(\ell) [3 \tilde{\alpha}(\ell) + 4 \tilde{\alpha}(\ellp)] }{[ \tilde{ \alpha}(\ell) + \tilde{\alpha}(\ellp)  ] ^{2}}.
\label{k2full}
\end{align}
For monodisperse sytems,  when $\alpha(\ell) = \alpha(\ellp) = \alpha $ and $\ell = \ellp = 1$, one recovers Odijk's scaling result $K_{2} = c_{0}^{2} K_{2}^{(11)} \sim 7c_{0} /24 \pi$ \cite{odijkelastic}.  From \eq{eqpitch}, we infer that  {the cholesteric pitch always decreases with overall rod concentration through  $p_{c} \sim q^{-1} \propto c_{0}^{-1} $}, irrespective of polydispersity. The length distribution will, of course, have an effect on the pitch,  but only through modification of the twist elasticity, as evident from \eq{k2full}. We will explore this in more detail in the next Section.

\section{Results for Log-Normal and Schulz-Distributed Rod Lengths}

 \begin{figure*}
\centering
\includegraphics[width= 1.6 \columnwidth]{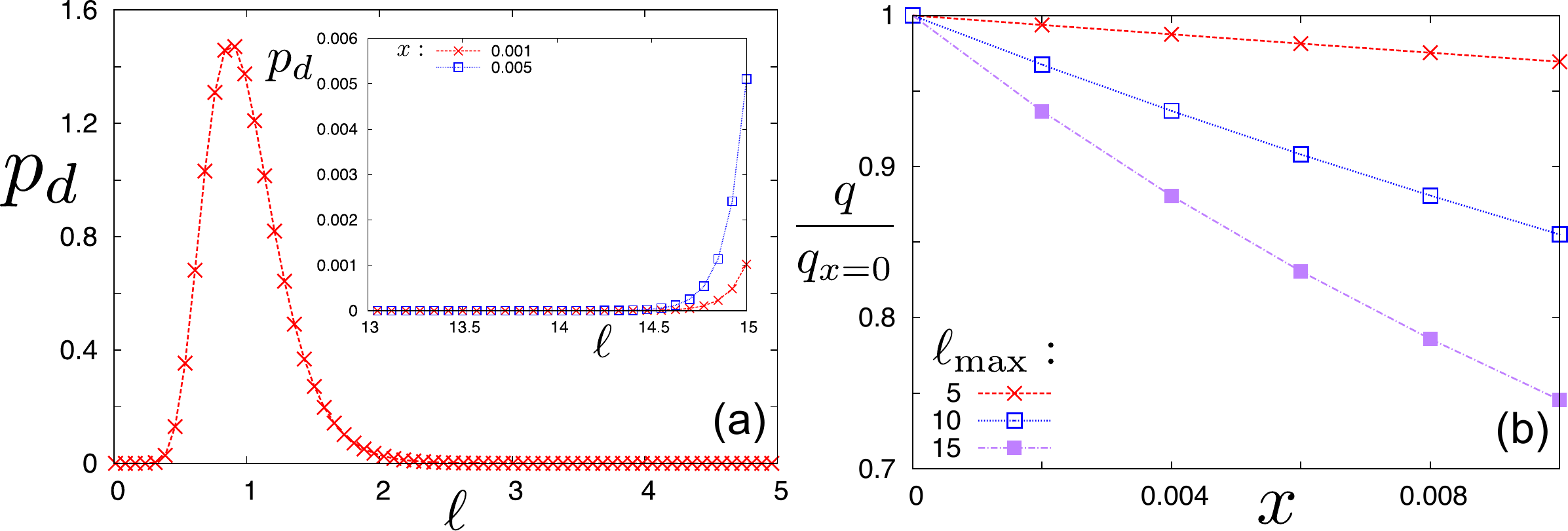}
\caption{ \label{fig3} (\textbf{a}) {Overview of a weakly bimodal log-normal distribution  ($\sigma = 0.3$) with an  exponential tail at $\ell_{\rm max} = 15$ at  different mole fractions  $x$ of long-rod dopants (inset)};  (\textbf{b}) Reduction of the cholesteric pitch wave number $q$, upon increasing the mole fraction $x$ of  large rods with $\ell_{\rm max}$ times the average rod length. }
\end{figure*}

A  typical size distribution {for polymers \cite{solc1975}, as well as for colloidal particles with quenched polydispersity \cite{Kiss_1999}, is the log-normal distribution, which is based on the logarithm of the rod length  following a normal distribution:}
\beq
p(\ell) = \frac{ 1}{(2 \pi )^{\frac{1}{2}} w \ell } \exp \left [ -\frac{( \ln \ell + \frac{w^{2}}{2} )^{2}}{2 w^{2}}  \right ],
\label{ln}
\eeq
with natural bounds $\ell_{\rm min} =0 $ and $\ell_{\rm max} \rightarrow \infty $. \eq{ln} has unity mean $\langle \ell \rangle = 1$, whereas the polydispersity $\sigma$ is connected to the standard deviation by:
\beq
\sigma = \left ( \frac{ \langle  \ell^{2} \rangle  - \langle \ell \rangle ^{2} } {\langle \ell \rangle^{2}} \right ) ^{\frac{1}{2}},
\eeq
 through $\sigma^{2} = e^{w^{2}} -1$. Finite-tail cutoffs lead to small corrections that are easily accounted for  numerically.
Alternatively, a commonly-used form representing polymer molecular weight distributions is the Schulz-Zimm  function \cite{schulz1939,zimm1948}:
\beq
p(\ell) = \frac{(1+z)^{1+z}}{\Gamma (1+z)} \ell^{z} \exp (-(z + 1) \ell ),
\eeq
which is normalized on the domain $0< \ell < \infty$ and has mean $\langle \ell \rangle = 1$ and polydispersity \mbox{$\sigma = (1+z)^{-1/2}$}.  The exponential tail renders cut-off effects far less serious than for the log-normal form~\cite{speranza_jcp2003}. The results for both distributions are shown in \fig{fig2}.  For the log-normal distribution cut-off values of $\ell_{\rm min} = 0.01$ and $\ell_{\rm max} = 20$ were used. The increase of the twist elastic modulus with polydispersity $\sigma$ appears significant and robust, as it is mostly insensitive to the shape of the length  distribution and the cut-off values. Clearly, introducing a spread of rod lengths at a given overall concentration induced a significant ``stiffening'' of the nematic fluid with respect to a twist distortion of the director.

\subsection{Effect of Large-Rod Dopants and Bimodality}

We will now explore the effect of doping a unimodally length-distributed nanorod cholesteric with a tiny fraction of large rods of length $\ell_{\rm max}$. Let us supplement the log-normal distribution \eq{ln} with a growing exponential tail to construct a weakly bimodal size distribution \cite{ferreiro2016}:
\beq
p_{d}(\ell) \sim p(\ell) +  x  e^{-a ( \ell_{\rm max} - \ell)}  \label{pd},
\eeq
 where $x \ll 1$ is the mole fraction of the added rods and $a  \gg 1$ quantifies the degree of bimodality.  \eq{pd} lacks a trivial normalization factor, which is included in the numerical calculations.  Moreover, the alteration of the log-normal parent distribution affects the renormalized average length, such that $\langle \ell \rangle = \int d \ell \ell p(\ell )  >1 $ which, in turn,  changes the helical amplitude $K_{t}$ through  \eq{ktll}. These effects are easily accounted for numerically.  Results for $a=10$ and a unimodal polydispersity $\sigma$ of 30\% (which seems a typical value, for example,  for CNCs \cite{lagerwall2014a}) are shown in \fig{fig3}, for different values of the maximum cut-off $\ell_{\rm max}$. The results demonstrate that adding even a very small fraction of long rods (less than 1\%) causes a significant reduction of the helical twist.  The strength of the reduction can be systematically tuned through the length of the doped rods. Since the bimodal twist reduction of the cholesteric system is imparted {mostly} through modification of its twist elasticity, the microscopic chiral properties of the doped rods are not imminently important provided their number fraction remains sufficiently small..

\subsection{Pitch Variation across the Isotropic-Cholesteric {Biphasic Region}}
 
  \begin{figure*}
\centering
\includegraphics[width=  1.5\columnwidth]{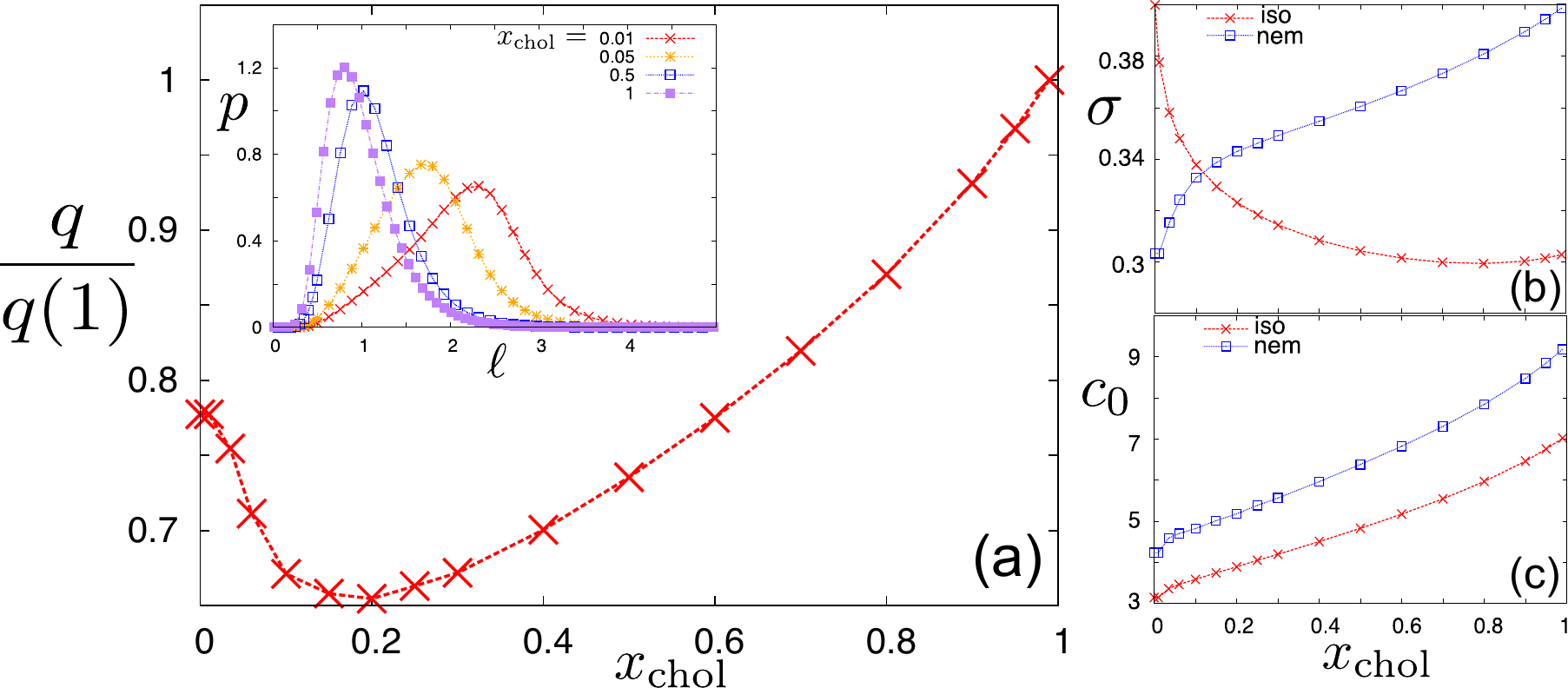}
\caption{ \label{fig4} (\textbf{a}) Variation of the pitch across the isotropic-cholesteric (I--C) {biphasic region}  for a nanorod system having a log-normal length distribution with $\sigma = 0.4$. Plotted is the pitch, $q$, renormalized to its value $q(1)$ at the C--I cloud point, versus the cholesteric  phase fraction, $x_{\rm chol}$.  The inset depicts a number of length distributions in the cholesteric phase at different phase fractions; The panels (\textbf{b}) and (\textbf{c}), on the right, indicate   the  polydispersity, $\sigma$,  and  the concentration, $c_{0}$, (\textbf{c}) of the coexisting phases versus $x_{\rm chol}$.   }
\end{figure*}

 We finish our analysis by investigating the behaviour of the cholesteric pitch within the isotropic-nematic {biphasic region}. The thermodynamics of phase transitions of length-polydisperse rod systems within the Gaussian Ansatz has been discussed in detail, in \cite{wensink_jcp2003}. We may determine coexistence between the isotropic and cholesteric phases by imposing equality of osmotic pressure and chemical potential, both of which are straightforward derivatives of the free energy \eq{free}.  At finite phase fractions, the distribution of rod lengths in each of the  coexisting phases is different from the imposed log-normal parent distribution. Concomitantly, the cholesteric pitch will depend non-trivially on the phase fraction, or the location within the {biphasic region}.  This is illustrated in \fig{fig4}, showing the variation of the pitch as well as the evolution  of the concentration and polydispersity of the cholesteric phase fraction across the biphasic region. Upon moving away from the isotropic-cholesteric (I--C) cloud point ($x_{\rm chol} = 0$), where only a infintesimal fraction of cholesteric phase has been formed (referred to as the ``shadow''),  the cholesteric unwinds initially and then rewinds (i.e., tighter pitch lengths)  close to the C--I cloud point  {($x_{\rm chol} = 1$)}. In the latter, where a negligible fraction of isotropic phase is left, the length distribution within the cholesteric phase equals the log-normal parental one. The non-monotonic trend of the pitch is not inflicted by the cholesteric concentration, which increases gradually upon $x_{\rm chol}$, but is the result of subtle changes in the length variation upon traversing the {biphasic region}. We remark that the polydispersity of the cholesteric phase is at its lowest  (about $\sigma  \approx 0.3 $) at the I--C cloud point, thus offering a simple means for purifying a polydisperse cholesteric system through successive sweeps of phase separation. In addition, splitting off the cholesteric phase fraction close to the I--C cloud point provides an effective way of ``filtering~out'' the largest rods,  given that the average rod length is larger than overall, as suggested by the distribution for $x_{\rm chol} = 0.01$ in the inset of \fig{fig4}.

\section{Conclusions}

Inspired by a recent upsurge in experimental studies on cholesteric self-organization of rigid chiral nanorods with quenched length polydispersity (most notably, microfibrils made of cellulose \cite{lagerwall2014a}, chitin \cite{REVOL1993,belamie2004}, and related biocomponents) we have extended the Onsager-Straley theory \cite{Onsager,straleychiral} for the cholesteric organization of  chiral rods with uniform length, towards the polydisperse case. The central assumptions underlying the theoretical analysis are the following: (i)  The rods are completely rigid and sufficiently slender, so as to respect the Onsager limit of infinite length-to-width ratio; (ii)  the local nematic alignment along the revolving director describing a twisted nematic is asymptotically strong, which justifies the use of a simple Gaussian variational approach \cite{odijkoverview} for the local orientational probability; and (iii) the helical deformation, $q$, of the director field is weak, on the scale of the average rod length $L_{0}$, so that $qL_{0}\gg 1$. We show that, with these criteria fulfilled, the Onsager-Straley theory can be cast in an algebraic form.   The determination of the pitch for a given length distribution requires relatively little computational cost, save for a straighforward numerical iterative procedure to determine the {length-dependence} of the variational parameter describing the degree of nematic order.  Our~main finding is that length polydispersity principally enhances the twist elasticity of a cholesteric material, with the helical twisting power (generated by the microscopic chirality of the rods) being only marginally affected. Quantitative examples are given of a pitch reduction generated by doping a polydisperse cholesteric system with long rods residing in the tail of the unimodal length distribution.. 

Without claiming to have presented an accurate theory for any  chiral nanorod assembly in particular, we believe the present algebraic  theory to be capable of providing a tractable and physically insightful tool that may  be helpful for interpreting and guiding experimental observations in these systems. In particular, our findings demonstrate that the isotropic-cholesteric phase transition can be used as a useful vehicle to purify or select chiral species of a certain length, or to fine-tune the pitch of polydisperse nanorod cholesterics.

\section{Gaussian Averages}

 The procedure for obtaining Gaussian averages needed for the computation of the twist elasticity $K_{2} ^{(\ell \ellp)}$ of a polydisperse nematic is given in Reference \cite{odijkelastic}. The following averages are required:
\begin{equation}
 \nonumber
  \langle \langle  | \gamma |^{3}   \theta^{2}   \rangle \rangle_{\psi_{G}}     \sim  3 \left ( \frac{\pi}{2} \right ) ^{\frac{1}{2}} \left ( \frac{1}{\alpha_{1} } + \frac{1}{\alpha_{2}} \right )^{\frac{1}{2}} \frac{2 \alpha_{1} + 5 \alpha_{2}}{\alpha_{1}^{2} \alpha_{2}} ,
   \end{equation}  
   \begin{equation}
  \langle \langle  | \gamma |   \theta^{4}   \rangle \rangle_{\psi_{G}}      \sim \left ( \frac{\pi}{2} \right ) ^{\frac{1}{2}} \left ( \frac{1}{\alpha_{1} } + \frac{1}{\alpha_{2}} \right )^{\frac{1}{2}} 
  \frac{8 \alpha_{1}^{2}  + 24  \alpha_{1} \alpha_{2} + 15 \alpha_{2}^{2}}{\alpha_{1}^{2} ( \alpha_{1} + \alpha_{2})^{2}} ,
 \end{equation}  
 \begin{equation}
 \nonumber
  \langle \langle  | \gamma |  \theta^{2} \theta^{\prime 2}   \rangle \rangle_{\psi_{G}}    \sim \left ( \frac{\pi}{2} \right ) ^{\frac{1}{2}} \left ( \frac{1}{\alpha_{1} } + \frac{1}{\alpha_{2}} \right )^{\frac{1}{2}} 
  \frac{6 \alpha_{1}^{2}  + 11  \alpha_{1} \alpha_{2} + 6 \alpha_{2}^{2}}{\alpha_{1} \alpha_{2} ( \alpha_{1} + \alpha_{2})^{2}}   , 
   \end{equation}
where we denote $\alpha_{1} = \alpha(\ell)$ and $\alpha_{2} = \alpha( \ellp ) $.  

\bibliographystyle{unsrt}
\bibliography{pub}

\end{document}